\documentclass[12pt]{article}
\usepackage{amsmath}
\usepackage{amssymb}
\usepackage{epsfig}
\usepackage{cite}
\usepackage{color,colordvi}

\usepackage[utf8]{inputenc}
\usepackage{amsmath}
\usepackage{amssymb}
\usepackage{graphicx}
\usepackage{tikz}

\begin{document}
\vspace{0.01cm}
\begin{center}
{\Large\bf Road Signs for UV-Completion } 

\end{center}

\vspace{0.1cm}

\begin{center}

{\bf Gia Dvali}$^{a,b,c,d}$\footnote{georgi.dvali@cern.ch}, {\bf Andre Franca}$^{a}$\footnote{andre.franca@physik.lmu.de}, {\bf Cesar Gomez}$^{a,e}$\footnote{cesar.gomez@uam.es}

\vspace{.6truecm}

{\em $^a$Arnold Sommerfeld Center for Theoretical Physics\\
Department f\"ur Physik, Ludwig-Maximilians-Universit\"at M\"unchen\\
Theresienstr.~37, 80333 M\"unchen, Germany}

{\em $^b$Max-Planck-Institut f\"ur Physik\\
F\"ohringer Ring 6, 80805 M\"unchen, Germany}

{\em $^c$CERN,
Theory Department\\
1211 Geneva 23, Switzerland}

{\em $^d$Center for Cosmology and Particle Physics\\
Department of Physics, New York University\\
4 Washington Place, New York, NY 10003, USA}

{\em $^e$
Instituto de F\'{\i}sica Te\'orica UAM-CSIC, C-XVI \\
Universidad Aut\'onoma de Madrid,
Cantoblanco, 28049 Madrid, Spain}\\

\end{center}

	\begin{abstract}
	\noindent  
 We confront the concepts of  Wilsonian UV-completion versus self-completion by {\it Classicalization} 
 in theories with derivatively-coupled scalars. 
  We observe that the information about the UV-completion road 
 is encoded in the sign of the derivative terms. 
   We note that the sign of the derivative couplings for which there is no consistent 
   Wilsonian UV-completion is the one that allows for consistent classicalons. 
   This is an indication that for such a sign the vertex must be treated as fundamental and the theory self-protects
   against potential inconsistencies, such as superluminality,  via self-completion by classicalization.   
   Applying this  reasoning to the UV-completion of the Standard Model, we see that the information 
   about the Higgs versus classicalization is encoded in the sign of the scattering amplitude of longitudinal W-bosons.  Negative sign excludes Higgs 
   or any other weakly-coupled Wilsonian physics.

	\end{abstract}
\thispagestyle{empty}
\clearpage

\section{Introduction}

 The Wilsonian paradigm of UV-completion is based on the existence of 
 weakly-coupled elementary degrees of freedom at arbitrarily short scales.  
 Of course, a given degree of freedom  (for example a pion) need not be a good description at all 
the scales and can become strongly coupled beyond certain cutoff (such as the QCD length),  where 
it has to be replaced by new weakly-coupled particles (such as quarks and gluons). 
  The crucial requirement however is that such a replacement must be  possible beyond an arbitrary cutoff. 
 
 According to this point of view the strong coupling appearing beyond some  perturbative cutoff  length $L_*$,  for example  in theories such as gravity or Nambu-Goldstone-type scalars,  is  
 an artifact of missing weakly-coupled degrees of freedom that must be integrated-in  in order 
 to restore the perturbative unitarity.  
  This approach has been extremely successful in theories 
 such as QCD and the electroweak interactions. 
 But, the question is, how far on the theory landscape it stretches. 
 
   Recently, an alternative concept  to non-Wilsonian self-completion was suggested in
   \cite{gia-cesar, class}. According to this view, a given theory may self-complete 
  without the need of new weakly-coupled elementary degrees of freedom beyond the cutoff 
 length.   Instead, their role is taken up by the collective excitations of multi-particle states
 composed out of soft {\it original}  quanta.   This concept was originally applied to  gravity in 
 \cite{gia-cesar}.  In \cite{class} it was generalized to other derivatively-interacting theories 
and was termed {\it classicalization}.  

 Classicalization is a deeply quantum-mechanical concept, and its essence is the following. 
Consider a Bosonic degree of freedom  $\phi$ with an effective  quartic  (self)coupling 
that  grows with the inverse wave-length of $\phi$-quanta as a certain power $n$.  The corresponding 
coupling constant $G$, sets the cutoff length (to be denoted by $L_*$) as 
\begin{equation} 
 L_*^{n} \, \equiv  \, \hbar \, G \, .
\label{lstar}
\end{equation}
Quantum-mechanically, the strength of the coupling of such quanta of a given wavelength $L$ is measured by the quantity, 
\begin{equation}
\alpha \, \equiv  \,  (L_*/L)^{n} \, \equiv  \, \hbar \, G \, L^{-n}.
\label{alpha}
\end{equation}
 Viewed in a Wilsonian context, by scattering the two quanta at a center of mass energy $E \, \gg \, \hbar/L_*$  
one probes distances $L \, \ll \, L_*$.  The quantum-mechanical coupling of such quanta (\ref{alpha}) is obviously strong, and thus the scattering violates perturbative unitarity. 
 According to the Wilsonian approach, the restoration of unitarity requires 
 integrating-in some 
 new weakly-coupled physics.  The idea of classicalization suggests a different route. 
   Instead of producing the two very hard quanta, the process is dominated by production of 
a state with many soft quanta, of wave-length  
\begin{equation}
 r_* \, = \, L_* (L_*E/\hbar) ^{{1 \over n -1}} 
 \label{rstar1}
 \end{equation}
 and with occupation number 
 \begin{equation}
 N \, = \,(Er_*)/\hbar \, .
 \label{N}
 \end{equation}
  Plugging $L \, = \, r_*$ in (\ref{alpha}) and comparing with (\ref{N}), we see that
 these quanta interact with the strength, 
 \begin{equation}
 \alpha \, = \, 1/N  
 \label{1overN}
 \end{equation}
 and thus are weakly-interacting for $N \, \gg \, 1$,  or equivalently for $E \, \gg \, L_*^{-1}$. 
 
 Thus, the essence of classicalization is to unitarize the high-energy scattering by replacing  $ 2 \rightarrow 2 $ hard scattering  by $ 2 \rightarrow N$ scattering into many  weakly-interacting soft quanta.  In a certain sense, the role of would-be UV-completing Wilsonian  degrees of freedom 
 is played by a collective weakly-coupled degree of freedom $N$. 
 However, $N$ is not a new degree of freedom, but a composite state of many soft and weakly interacting "old" quanta. 

   This multi-particle quantum state is referred to as a {\it clasicalon}. It is conceivable to assume that 
   a classicalon must have a well-defined classical limit.  What is this limit? 
 For understanding this notice that the (semi)classical limit is achieved by taking, 
 \begin{equation}
 N \, \rightarrow \, \infty \, ~~~\, L_* \, \rightarrow \, 0 \,~~~ r_* \, = \, {\rm fixed} \, .
 \label{limit}
 \end{equation}
 In addition, depending on whether we wish to reach semi-classical or classical limit, we 
 either keep $\hbar$ fixed or send it to zero.     
     
      It is clear that if a well-defined classical limit exists for the classicalon state, it should correspond to a configuration that is a solution of the classical equations of motion, with $r_*$ being its 
      characteristic integration constant.   The most celebrated solution 
 of this type is the Schwarzschild black hole in gravity, with $r_*$ being the Schwarzschild radius. 
 The above large-$N$ quantum portrait of black holes was systematically developed 
 in \cite{NQuantum}. In the present paper we shall be interested in applying it to 
 spin-0 theories.   Hence we observe the following general quantum-to-classical dictionary. 
 
   The $r_*$ radius that quantum mechanically defines the characteristic wave-length of the 
 multi-particle state produced at energy $E$, classically corresponds to a geometric 
 radius of a static field-configuration of the same energy $E$.         
     The necessary condition for classicalization is the growth of the $r_*$-radius with energy.\footnote{It was suggested \cite{classsafe} that the marginal case, for which $r_*$ freezes, could serve as a connecting point between classicalization and Wilsonian asymptotic safety. This interesting 
point will not be discussed here. We shall focus on the cases of strong classicalization, 
when $r_*(E)$ grows with $E$ as a power law.} 
  
 In the present paper we would like to address the following question. 
 For a given low energy theory, how does nature decide which road to UV-completion
 (Wilsonian versus Classicalization) to take?  This question was partially answered in \cite{class} and 
 \cite{gia}.   In \cite{class} is was shown that in a completion by a linear sigma-model the static classicalons no longer exist because the phase 
 gradients back reacts at the Higgs mode.   
 In \cite{gia} the issue was analyzed in more general effective field theoretic terms and it was shown that the two concepts are mutually-exclusive. 
 By analyzing the time-dependent evolution of the wave-packets it was shown that any potential softening of the derivative vertex by 
integrating-in a new weakly-coupled physics (that could 
 restore perturbative unitarity in  $2\rightarrow 2$ scattering) automatically de-classicalizes the theory by collapsing  $r_*$ radius to distances below $\hbar /E$.
 
 To set the road signs towards UV-completion we will consider spin-zero field theories that at tree-level produce scattering amplitudes growing with energy as $c(sL_*^2)^2$. The problem of UV completion is equivalent to the problem of how to unitarize those theories at energies larger than the unitarity bound $L_*^{-1}$. As discussed above there are two possible options: Wilsonian unitarization and Classicalization. In the first case the Lagrangian suffering unitarity problems is interpreted as the effective low-energy Lagrangian resulting from 
 integrating-out some extra degrees of freedom in a more basic and 
 UV-complete theory. In this case the tree level interaction leading to the non unitary growth is not fundamental and results from the integrating-out procedure that uniquely sets the sign of $c$ to be positive. The other possibility is that the theory possesses a non perturbative spectrum of finite energy configurations, the classicalons,  with energies bigger than the unitarity bound $L_*^{-1}$ and composed of large number $N$ of the quanta of the theory. In this case the tree level vertex leading perturbatively to a non unitary growth should be considered as {\it fundamental}. Then,  unitarization takes place not by resolving this vertex, but by classicalon dominance of the amplitude at high energies. In this case the sign of $c$ is not determined by the existence of extra weakly-interacting UV degrees of freedom, but by the existence of the non perturbative spectrum of classicalons\cite{class}. Nicely enough this sets the sign of $c$ to be minus i.e. the opposite to the sign implied by the Wilsonian completion. 
  Notice,  that the sign-sensitivity to the UV-completion is also hinted by the very different time evolutions of the system in the two cases\cite{tetradis}.  
 
 The previous result immediately leads to a caveat since classicalons exist precisely for the sign that are known \cite{nima} to lead to superluminal propagation of modes on certain backgrounds.  However, superluminality can be a disaster for Wilsonian UV-completion, but not necessarily 
so for classicalization, since the latter limits the possible momentum-transfers and thus the boost-factors relative to the would-be problematic background.  
This could suggest that classicalization offers some sort of a self-protection against superluminality. 

A nice example where we can apply our findings is in the frame of the recent derivation of the $a$ theorem\cite{komar}. In that case the auxiliar dilaton lagrangian contains at tree level a vertex producing amplitudes $a(sL_*^2)^2$ with $a = a(UV)-a(IR)$ and $L_*$ the scale of the corresponding spontaneous breakdown of the conformal symmetry. Thus in this case the positive sign is equivalent to the $a$-theorem. Moreover the sign consistent with the $a$-theorem is the one corresponding to UV-completion of the dilaton lagrangian in Wilsonian terms, something that is guarantee by the existence of a UV CFT fixed point. 

Finally,  an important phenomenological consequence of our results is that it allows us to read from the sign of the scattering amplitude for longitudinal $W$-bosons at low energies the way the Standard Model unitarizes. If the sign is positive, we can have Higgs or any other form of Wilsonian UV-completion (e.g., technicolor).  But, if the sign is negative, the theory will reveal us a different form of unitarization by classicalization.   The detection of the sign can provide an important cross-check for the Wilsonian nature of the
potentially-discovered unitarizing physics.

\section{Classicalons versus Wilsonian UV-Completion}

\subsection{Goldstone Lagrangian}

Let's start with the following effective theory of a single real scalar field, which we shall refer as the Goldstone Lagrangian:

\begin{equation}
{\cal L}=\frac{1}{2}\left(\partial_{\mu}\phi\right)^{2}\, + \epsilon \, \frac{L^{4}_*}{4}\left(\partial_{\mu}\phi\right)^{4}\, ,\label{eq:goldstone}
\end{equation}
where the parameter $\epsilon$ can take the value $\pm 1$.  
Static spherically-symmetric  field configurations in the above system  
were studied in \cite{nonlinearvector} where Lagrangian (\ref{eq:goldstone})
was obtained as a decoupling limit of a massive non-linearly interacting 
vector field.  This scalar model was later introduced in 
\cite{class} as a simple prototype for classicalizing system. 
As it was shown there,  this system admits static (singular) spherically-symmetric solutions, {\it classicalons}, which satisfy the equation of motion
\begin{equation}
\partial^{\mu}\left\{ \partial_{\mu}\phi+ \epsilon L_{*}^{4}\partial_{\mu}\phi\left(\partial_{\nu}\phi\right)^{2}\right\} =0 \, .\label{eq:eomgoldstone}
\end{equation}
In order to see this, we can look for static spherically symmetric solutions in the vacuum. These should satisfy the cubic algebraic equation on $\partial_{r}\phi$
\begin{equation}
\partial_{r}\phi\left(1-\epsilon L_{*}^{4}\left(\partial_{r}\phi\right)^{2}\right)=\frac{ML_{*}}{r^{2}}\label{eq:goldstonestatic}
\end{equation}
where $r$ is a radial coordinate and where we have introduced $ML_{*}$ as an integration constant.

We can identify two distinct regimes. For $r\rightarrow\infty$, the contribution from the nonlinearities are irrelevant and we have the linear solution
\begin{equation}
\phi(r\rightarrow\infty)\sim\frac{r_{*}^{2}}{L_{*}^{2}}\frac{1}{r}
\end{equation}
while for $r\rightarrow0$, we have
\begin{equation}
\phi(r\rightarrow0)\sim\frac{r_{*}}{L_{*}^2}\left(\frac{r}{r_{*}}\right)^{1/3}
\end{equation}
and $r_{*}$ is defined as the scale at which the two 
solutions become comparable, 
\begin{equation}
r_{*}=\left(ML_{*}\right)^{1/2}L_{*} \, .
\end{equation}

\begin{figure}[]
\centering{}
\includegraphics[scale=0.7]{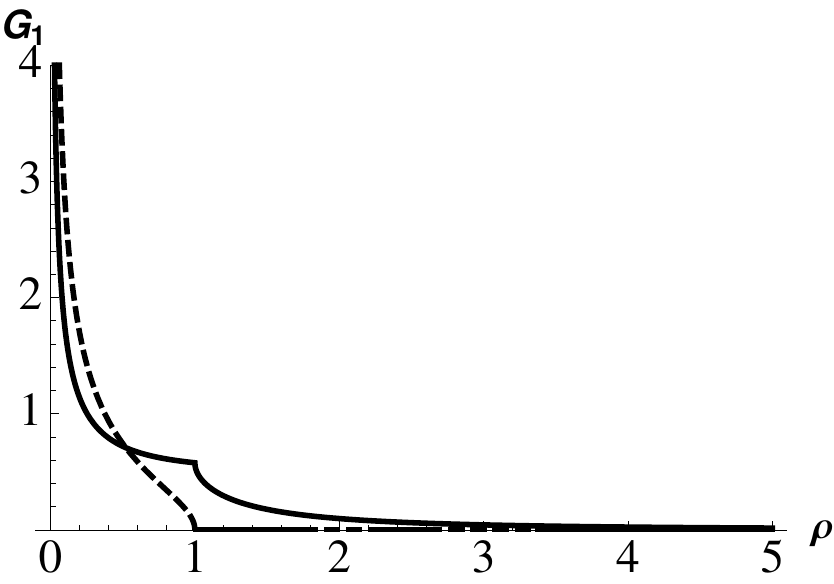}
\includegraphics[scale=0.7]{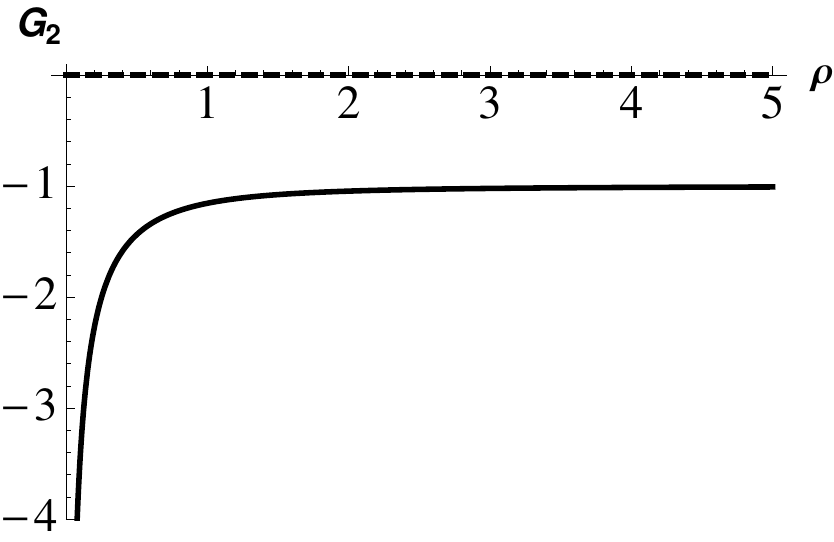}
\par
\centering{} 
\includegraphics[scale=0.7]{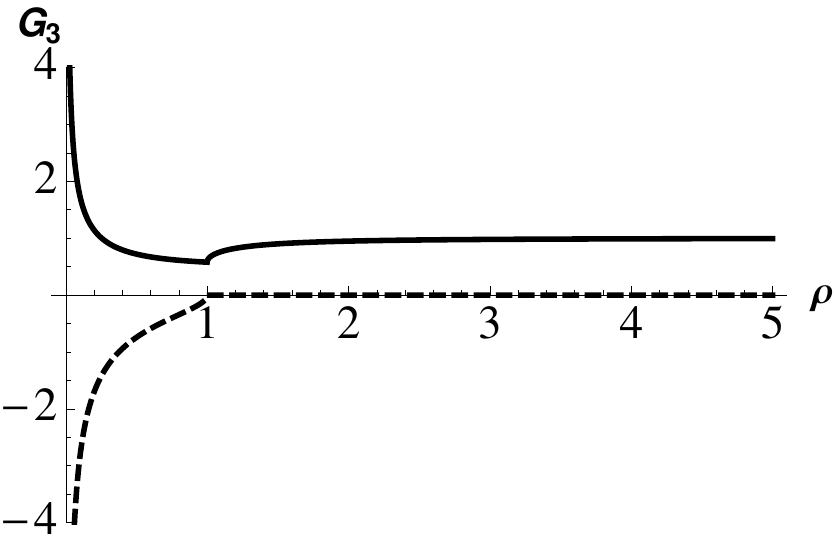} 
\caption{Solutions of (\ref{eq:goldstonestatic}) with $\epsilon = +1$ for $\partial_{r}\phi$, where the thick line is the real part and the dashed line is the imaginary part of the solution.}
\label{fig:roots}
\centering{}
$ $

$ $
\includegraphics[scale=0.8]{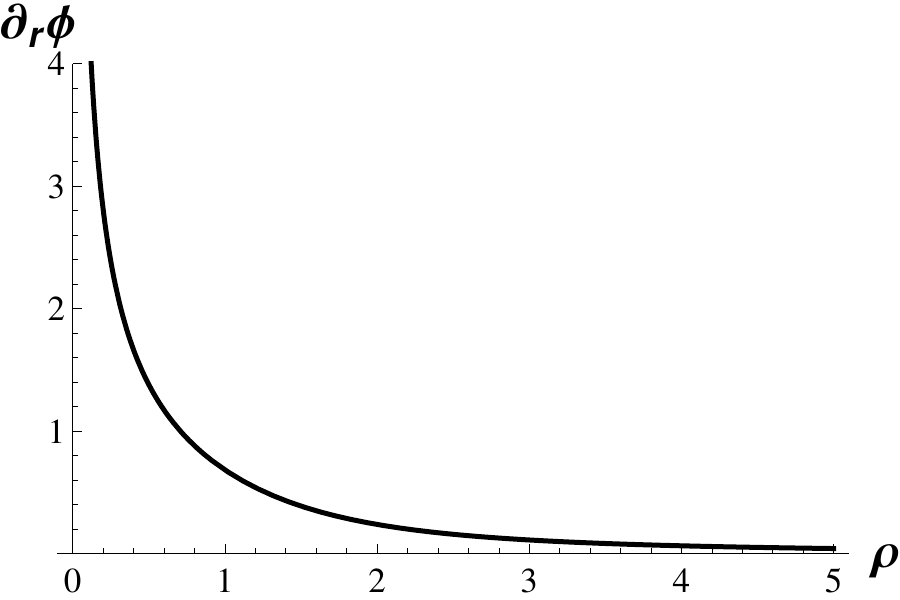}
\caption{Classicalon solution for $\epsilon = -1$.}
\label{fig:minus}
\end{figure}

 The important point is that the continuous classicalon solution exists only for  $\epsilon \, = \,  -1$ \cite{class, tetradis}. To show that this is the case, we can analitically solve the algebraic cubic equation (\ref{eq:goldstonestatic}) for $\partial_{r} \phi$. 
Under the rescalings 
$r \rightarrow\rho=\frac{r}{r_{*}}$, 
$\phi \rightarrow\frac{L_{*}^2}{r_{*}}\phi$, 
we have the roots of (\ref{eq:goldstonestatic}) shown in Fig (\ref{fig:roots}), where it can be seen that for $\epsilon \, = \, +1$ the only solution that remains real is unphysical, as it diverges as $\partial_{r}\phi\rightarrow L_{*}^{-2}$ for $r\rightarrow \infty$.

For $\epsilon \, = \, -1$, we have the static configuration whose behavior for $\partial_{r} \phi$ is shown in Fig (\ref{fig:minus}).  This  is a well-defined classicalon configuration.  
As we shall see, this is the sign for which no sensible Wilsonian UV-completion exists.  In the opposite case, $\epsilon = + 1$, the Wilsonian UV-completion exists, but classicalon solution is not well defined everywhere. 

We can also investigate what happens if we supplement this theory with higher order operators. Suppose that we take $\epsilon = -1$ and add a $\left(\partial\phi\right)^{2n}$ vertex suppressed by some scale $\Lambda$ that may or may not coincide with $L_{*}^{-1}$, 
\begin{equation}
{\cal L}=\frac{1}{2}\left(\partial_{\mu}\phi\right)^{2}\, - \frac{L^{4}_*}{4}\left(\partial_{\mu}\phi\right)^{4} \, + \frac{\epsilon_n}{\Lambda^{4n-4}}\left(\partial_{\mu}\phi\right)^{2n} \, .
\end{equation}
Even if it is still possible, for $\epsilon_n > 0$, to unitarize the $\left(\partial\phi\right)^{2n}$ vertex above the scale $\Lambda$ by integrating-in some weakly coupled physics, this will only tame the growth of the scattering amplitude for processes with $2n$ or more external legs. This operator -- and whatever physics it is embedded into -- cannot help with the unitarization of the $2\rightarrow 2$ amplitude coming from the quartic term.   Thus, regardless 
of higher order operators, for $\epsilon \, = \, -1$, the theory is not expected 
to have any weakly-coupled Wilsonian UV-completion and the only chance 
is to self-complete  by classicalization. \footnote{
Generalizing the previous analysis, in case $\epsilon_n=(-1)^n$, then we can be sure that a classicalon solution exists regardless of whether or not this higher order vertex is unitarized by weakly coupled physics. In case the theory choses to embed this term into Wilsonian physics, then as long as the scale $m$ of this new weakly coupled physics is much lower than the unitarity-violating scale, $m\ll\Lambda$,  the contribution from this vertex will always be subdominant with respect to the classicalon background, and we still can expect the theory to classicalize.} 

In order to understand the correlation between the sign of $\epsilon$ and the 
nature of UV-completion, it is instructive to complete the theory by embedding 
it into a weakly-coupled linear sigma model,  
\begin{equation}
\mathcal{L}=|\partial_{\mu}\Phi|^{2}-\frac{\lambda}{8}(2|\Phi|^{2}-v^{2})^{2} \, , 
\label{eq:higgs}
\end{equation}
where $\Phi$ is a complex scalar field carrying two real degrees of freedom, and the Lagrangian is invariant under the global $U(1)$ transformation
\begin{equation}
\Phi\rightarrow e^{i\theta}\Phi \, .
\label{eq:u1}
\end{equation}
In the ground state of the theory, the symmetry (\ref{eq:u1}) is spontaneously broken by the vacuum expectation value of $\Phi$, $\left\langle \Phi^{*}\Phi \right\rangle =\frac{1}{2}v^{2}$, so we can parametrize the degrees of freedom as
\[
\Phi=\frac{1}{\sqrt{2}}(v+\rho)e^{i\phi/v} \, , 
\]
in terms of a radial (Higgs) mode  $\rho$ and a Goldstone boson $\phi$,  which under spontaneously broken $U(1)$ transforms as,  
\begin{equation}
{\phi \over v} \,  \rightarrow \, {\phi \over v} \, + \, \theta \, .
\end{equation}

In terms of the fields $\rho$ and $\phi$ the Lagrangian becomes, 
\begin{equation}
\mathcal{L}=\frac{1}{2}(\partial_{\mu}\rho)^{2}+\frac{1}{2}\left(1+\frac{\rho}{v}\right)^{2}(\partial_{\mu}\phi)^{2}-\frac{\lambda^{2}}{8}(\rho^{2}+2\rho v)^{2} \, ,
\end{equation}
from where $\rho$ acquires a mass of $m=\lambda v$.

For energies $E\ll\lambda v$, we can integrate out  $\rho$ and write down an effective low energy theory for $\phi$.
Integrating out $\rho$ through its linear  order equation,  
\begin{equation}
\left\{ \square+m^{2}\right\} \rho=\frac{1}{v}(\partial_{\mu}\phi)^{2} \, , 
\end{equation}
we get the following leading order low energy effective equation for 
$\phi$,  
\begin{equation}
\partial^{\mu}\left\{ \partial_{\mu}\phi\left(1+L_{*}^{4}\frac{m^{2}}{\square+m^{2}}(\partial_{\nu}\phi)^{2}\right)\right\} =0
\end{equation}
where we have defined a cutoff length $L_{*}^{2}=\sqrt{2}(mv)^{-1}$. 
 In the low energy limit, $m^2 \gg \square$, this recovers (\ref{eq:eomgoldstone}) with $\epsilon = 1$.  

 The origin of the positive sign is now clear. 
Notice  that if we would want to obtain the similar low energy effective Lagrangian but with $\epsilon = -1$,  we had to flip the sign of the $m^2$ term in Lagrangian 
of $\rho$. This would mean that we had integrated out a {\it tachyon}  as opposed to a normal particle,  which makes no sense.  We thus see that there is no sensible weakly-coupled linear sigma model 
that in the low energy limit gives the Lagrangian admitting classicalon solutions. 
 It can be checked explicitly \cite{class} that the $r_*$ radius of static classical sources collapses when the Goldstone theory is embedded in the linear sigma model.
 This effect can be understood as a particular manifestation of a very general {\it de-classicalization} phenomena by weakly-coupled UV-completing physics \cite{gia} that we shall discuss in more details below.  
 The above discussion shows why the classicalizing theory cannot be obtained as a low energy limit of a weakly-coupled UV-completion. 

\subsection{DBI}

 Another example that illustrates incompatibility between the classicalons and UV-completion by a weakly-coupled theory, is provided 
 by  the embedding of the Goldstone model (\ref{eq:goldstone}) into 
 a Dirac-Born-Infeld (DBI) theory. 
 The Lagrangian (\ref{eq:goldstone}) can be thought as an expansion of 
 DBI type Langangian
 \begin{equation}
 {\cal L}_{DBI} \, = \, \epsilon_1 L_*^{-4} \sqrt{1 \, + \, \epsilon_2 \, L_*^4 \, \left(\partial_{\mu}\phi\right)^{2}}, 
 \label{DBI}
 \end{equation}
 where parameters $\epsilon_{1,2}$ can take values $\pm 1$. 
 For the values $\epsilon_1 \, = \, \epsilon_2 \, = \, 1$ this theory admits a classicalon solution \cite{class}, 
\begin{equation}
\partial_r \phi \, = \, {r_*^2 \over L_*^2} {1 \over \sqrt{ r^4 \, + \, r_*^4} }   \, ,  
\label{DBIclass}  
\end{equation}
where $r_*$ is an integration constant.   
On the other hand, the sensible embedding into a weakly-coupled theory
is only possible for $\epsilon_1 = \epsilon_2 = -1$. 
To see this, note that the action (\ref{DBI}) can be viewed as an effective low energy action
(Nambu-type action) describing the embedding of a $3$-brane (domain wall) in a five-dimensional space-time, with $\phi$ being a Nambu-Goldstone mode of spontaneously broken translational 
invariance.  Expanding this action in powers of $\phi$ we get the action 
very similar to (\ref{eq:goldstone}),  
\begin{equation}
{\cal L}=\epsilon_1 L_*^{-4} \, + \, \epsilon_1\epsilon_2 \, \frac{1}{2}\left(\partial_{\mu}\phi\right)^{2}\, - \epsilon_1 \epsilon_2^2 \, \frac{L^{4}_*}{4}\left(\partial_{\mu}\phi\right)^{4}
 \, + \, ... \,  . 
 \label{braneaction}
\end{equation}
 The first term represents the brane tension with negative sign and this fixes $\epsilon_1 = -1$. 
 The positivity of the kinetic term fixes $\epsilon_2 = -1$. As a result the sign of the 
 next term is fixed to be the one that does not admit classicalon solutions.  
 Having a classicalizing  theory requires $\epsilon_1 = +1$, which would 
 give a wrong sign brane tension.  We thus see that,  just like in the Higgs case, 
 the  weakly-coupled UV-completion is possible for the sign that does not admit the classicalon solutions and vice-versa.

\subsection{Evidence from Spectral Representation}

Impossibility of sensible weakly-coupled  UV-completion for $\epsilon \, = \, -1$ 
 can be seen from the following general argument.
The existence of such UV-completion would imply that the effective four-derivative vertex is a result of integrating-out some 
weakly-interacting physics that couples to $\phi$ in form of an effective current 
\begin{equation}
\partial_{\mu}\phi \partial_{\nu} \phi \, J^{\mu\nu} \, ,   
\label{current}
\end{equation}
where $J_{\mu\nu}$ is some effective (in general composite) operator that 
encodes information about the given UV-completing physics.  
 From the symmetry properties it is clear that the current  $J_{\mu\nu}$ can transform either as  spin-$2$ or spin-$0$ under the Poincare group. 
 The effective  four-derivative vertex of $\phi$ is then result of a non-trivial  $\langle J_{\mu\nu} J_{\alpha\beta} \rangle$ correlator. 
 The positivity of $\epsilon$ follows from the  positivity of the spectral function in the K\"allen-Lehmann  spectral representation 
 of this correlator. 
The most general ghost and tachyon-free spectral representation of this current-current correlator is (see, e.g., \cite{spectral}), 
\begin{eqnarray}
 \label{spectral}
 \langle  J_{\mu\nu} \, J_{\alpha\beta} \rangle \,  &=& 
 \, \int_0^\infty\, dm^2 \, \rho_2
(m^2)\frac{{1\over 2} (\tilde{\eta}_{\mu\alpha} \tilde{\eta}_{\nu\beta} \, +  \, \tilde{\eta}_{\mu\beta} \tilde{\eta}_{\nu\alpha}) \, - \, \frac{1}{3}\tilde{\eta}_{\mu\nu} \tilde{\eta}_{\alpha\beta}}{\square \, + \, m^2} \,  + \, \nonumber \\
 &+& \, \int_0^\infty  \, dm^2 \, \rho_0(m^2)
\frac{\eta_{\mu\nu} \eta_{\alpha\beta}}{\square \, +\, m^2} \, , 
\end{eqnarray}
where $\tilde{\eta}_{\mu\nu} \, = \, \eta_{\mu\nu} \, + \, {\partial_{\mu}\partial_{\nu} \over m^2}$ 
and $\rho_2(m)$ and $\rho_0(m)$ are the spectral functions corresponding to massive spin-$2$ and spin-$0$ poles respectively.  
 The crucial point is that  the absence of ghost and tachyonic poles demands that  these spectral functions are strictly positive-definite and 
 vanish for $m^2 \, < \, 0$ (the latter condition fixes the lower bound of integration). The entire weakly-coupled UV-dynamics is encoded in the  detailed form of these spectral functions,  which is completely unimportant for us except for the signs.  
 
 Convoluting this expression  with $\partial_{\mu}\phi \partial_{\nu}\phi$,  and ignoring high-derivatives,  it is obvious that  
 the coefficient of an effective low energy vertex is strictly positive, 
 \begin{equation}
 \label{lowlimit}
\partial_{\mu} \phi \partial_{\nu} \phi \langle J^{\mu\nu} J^{\alpha\beta} \rangle \partial_{\alpha}\phi \partial_{\beta}\phi  \, 
\rightarrow  \, 
(\partial_{\mu} \phi \partial^{\mu} \phi )^2   \, \int_0^{\infty} \, dm^2 \, \left({2 \over 3} {\rho_2(m) \over m^2}  \, + \, {\rho_0(m) \over m^2} \right ) \, .
\end{equation}
It is pretty clear  that having a negative sign for the coefficient requires either a ghost or a tachyonic pole. 
 The Higgs-completion  example considered in the previous section corresponds to a particular choice $\rho_2(m^2)\, = \, 0$ 
 and $\rho_0 (m^2) \, \propto  \, \delta(m^2 \, - \,  (\lambda v)^2)$.    
 
  Thus, the negative sign cannot be obtained by integrating-out any weakly-coupled Wilsonian physics. 
  However, instead of dismissing such a possibility, we should take 
this as a message that the theory tells us that we have to abandon the Wilsonian view, and treat the quartic vertex as {\it fundamental}.  The road that the theory chooses in such a case is  UV-completion through classicalization.

\section{Classicalization and Superluminality}

    We observed that in the Goldstone example the static classicalon solutions 
    are present precisely for the sign of the derivative interaction that as observed in \cite{nima}
    leads to backgrounds with superluminal propagation.  Such a superluminal propagation 
    usually would be a disaster in theories with  Wilsonian weakly coupled UV-completion, 
   but no such completion exists for the given sign.  Instead, the presence of classicalons is a signal that the theory chooses to self-complete by classicalization.  For such a completion 
 superluminality need not imply violation of causality. 
    To explain the reason, let us first reproduce the argument why superluminality appears 
 and why this may lead to a problem.  Consider  the goldstone Lagrangian (\ref{eq:goldstone}). 
 In this theory one can consider an extended field configuration that locally has a form 
 $\phi_{cl} \, = \, c_{\alpha} x^{\alpha}$,  where $c_{\alpha}$ are constants that are chosen 
 to be sufficiently small, so that the invariant is well-below the cutoff scale, 
 $ \left(\partial_{\mu}\phi\right)^{2} \,= \, c_{\alpha} c^{\alpha} \,  \ll \, L_*^{-4}$. 
 On such a background the linearized perturbations  $\phi \, = \, \phi_{cl} \, + \, \delta \phi$
 sees an effective Lorentz-violating metric,   
   \begin{equation}
 \eta_{\mu\nu} \, + \, \epsilon \, L_*^4 \, c_{\mu}c_{\nu}  \, + \, ... \, , 
 \label{metric} 
 \end{equation}
  which gives a superluminal dispersion relation for $\epsilon\, < \, 0$. 
In order  for this superluminality to become an inconsistency, one should be able to create 
  closed time-like curves and to send signals into the past.  
  Such a situation can be arranged by a set of highly boosted observers. 
  However, here we reach a subtle point.  In order to send a signal into the past at least some 
  of the observers must be boosted relative to the background with trans-cutoff center of mass energies. 
  Such a boost relative to a background is {\it not} a symmetry transformation and is physical. 
  So to rely on such a thought experiment we have to be sure that the interaction between 
  an observer and the background allows for such a boost.  Here comes the issue of the UV-completion. Since the center of mass energies are trans-cutoff, the legitimacy of the boost 
  depends on the UV-completion.  If the UV-completion is by Wilsonian weakly-coupled physics 
  (which is an implicit assumption of ref\cite{nima}), then boosts are allowed, since 
  for such UV-completions the cross-sections diminish at high energies, and a background is 
  not an obstacle for the boost.  However, as we have seen,  for the superluminal sign the Wilsonian UV-completion is absent anyway.  Instead, the theory allows classicalons,  which is  
 an indication that the theory chooses the classicalization path for UV-completion. 
  In such a case, the trans-cutoff boosts are a problem, since the cross-section increases with energy and any attempt to boost 
  an observer relative to the background with trans-cutoff energy 
  per-particle should result into creation of many soft quanta that will cutoff 
  the boost.  In this way, the system is expected to self-protect against creation 
  of closed time-like curves and violation of causality.
  
    Interestingly, in \cite{cron} it was argued that some sort of "chronology protection" may exist in a class of derivatively-coupled theories.  This work does not establish the connection with the nature of UV-completion, which is central 
 to our discussion. Nevertheless,  it can be viewed as an indirect  supporting evidence for the message we are trying to bring across: The nature of UV-completion is central to for consistency of theories admitting the superluminal backgrounds.  The general idea is simple. The classical backgrounds are IR effect, but they can play the crucial role in UV-dynamics 
in theories that in deep-UV become IR! Such are the classicalizing theories.   
  This is why the understanding of UV-completion becomes decisive for understanding the potential problems with superluminality.

 \section{Classicalization and the $a$-theorem}
 The $a$-theorem is the four dimensional generalization of Zamolodchikov's $c$-theorem in two dimensions \cite{zamo}. What the theorem establishes is that for a RG flow between two CFT fixed points , at the UV and the IR respectively, Cardy's $a$-function \cite{cardy} fulfills  the strong inequality $a(UV)>a(IR)$. Since along the flow the theory is not conformal,  the t'Hooft's anomaly matching conditions should be appropriately improved. This has been recently done in \cite{komar} based on previous results in \cite{stheisen}. The key idea is to interpret the theory along the flow as a spontaneously broken CFT with the dilaton as the corresponding Nambu-Goldstone boson. More precisely, given the UV CFT we add a relevant deformation to induce the flow as well as the coupling to the dilaton field. This is done in order to restore the conformal invariance leading to a total $T_{\mu}^{\mu}=0$. The vacuum expectation value (VEV)  $f$ setting the breaking of the conformal symmetry defines the decay constant of the dilaton field. Along the flow some massless UV degrees of freedom will become massive. The IR fixed point is obtained after integrating these out. Thus,  the final theory in the IR contains in addition to the IR CFT the low energy effective theory for the dilaton. This effective theory is
  \begin{equation}\label{one}
 \mathcal{L} \, = \, (\partial \phi)^2 \,  + \,  2a \,  {1 \over f^4}\,  (\partial \phi)^4 \, , 
 \end{equation}
 with $a$ satisfying the anomaly matching condition: $a(UV)-a(IR) =a$. Thus,  the $a$-theorem follows from the sign of the derivative-coupling of the effective low energy theory for the dilaton. 
 
 The connection with classicalization is now pretty clear. In fact the effective low energy theory for the dilaton is, as we have discussed in the previous sections, of the type of theories that, {\it depending on the sign of the derivative coupling}, can be self-completed in the UV by classicalization. 
 
 In order to understand the meaning of the $a$-theorem let us focus on the effective Lagrangian (\ref{one}). By itself this theory has a unitarity bound at energies of order $f$. This is obvious from the scattering amplitude that scales like
 \begin{equation}\label{two}
 A(s)\sim \frac{2as^2}{f^4} \, .
 \end{equation}
 In order to make sense of this theory we need to complete it at energies $E>f$. In the previous setup it is obvious how the effective theory of the dilaton is UV completed. Namely,  the completion takes place by the UV degrees of freedom of the UV CFT fixed point we have started with. In other words, the effective theory (\ref{one}) is, by construction, completed in the UV $E>f$ in a Wilsonian sense. The interesting thing is that this Wilsonian completion determines the sign of $a$ to be positive and therefore the proof of the $a$-theorem. This result directly follows from our previous discussion in the sense that for a negative sign of $a$ the theory cannot be completed in Wilsonian sense. 
 
 In other words, the sign of the derivative coupling -- and therefore the $a$-theorem -- depends on how the theory tames the growth of the amplitude (\ref{two}), i.e., on how the low energy effective theory of the dilaton unitarizes. A Wilsonian unitarization forces this sign to be positive. Therefore,  once we embed the dilaton dynamics in a flow with a well-defined UV CFT fixed point, the sign is forced to be positive, leading to the $a$-theorem. We cannot reach this conclusion directly from the Lagrangian (\ref{one}). In fact, general arguments based on dispersion relations for the dilaton scattering amplitude necessarily hide the key assumption on how the growth of the amplitude at high energies has been tamed. 
 
 Classicalization tells us however what is the physics when the sign is negative. In this case the theory unitarizes at high energies by classicalization. This means that the scale $f$ setting the unitarity bound becomes the limit on 
 length-resolution in the sense that the theory at higher UV energies turns 
 into a theory that probes IR scales. In particular, we can suggest the following conjecture. Let us start with a CFT in the IR and let us add an {\it irrelevant} operator and a coupling to a dilaton in order to keep the conformal invariance. If the effective theory of this dilaton has negative sign for the derivative self-coupling, the theory will classicalize in the UV.

 \section{Implication for UV-completion of the Standard Model}
   
    Our observations have important implication for determining the  UV-completion of the Standard Model. 
 If the scattering of the longitudinal $W$-s is unitarized by Higgs or any other weakly-coupled Wilsonian physics, the sign 
 of the four-derivative self-coupling must be positive. In the opposite case no weakly-coupled Wilsonian UV-completion is possible, but the theory 
 makes up due to the existence of classicalons, indicating that unitarization happens through classicalization.  
     This is remarkable, since  measurement of the sign of the  longitudinal  $WW$-scattering amplitude  can give a decisive information 
     about the UV-completion of the theory.  
   
   For completeness, let us repeat our arguments for the non-abelian case.  Since we are concerned with the scattering of the longitudinal $W$-s, 
   which are equivalent to Goldstone bosons, we shall work in the gaugeless  limit. 
    
     The Standard Model Lagrangian (with Higgs)  then reduces to  a Nambu-Goldstone model with spontaneously broken  $SU(2)$ global symmetry,   
 \begin{equation}
 \partial_{\mu} H^{a*} \partial^{\mu} H_a\,   -  {\lambda^2 \over 2} \, \left(H^{a*}H_a \, - \, \frac{v^2}{2}\right)^2 \,  ,
  \label{ng}
  \end{equation}
where $H_a\, (a=1,2)$ is an $SU(2)$-doublet scalar field. 

   Following \cite{class},  we shall now  represent the doublet  field in terms of the radial (Higgs) and Goldstone degrees of freedom,  
$H_a \, = \,  U_a(x) \,\rho(x)/\sqrt{2} \, = \, 
 ( cos\theta e^{i\alpha}, \, -sin\theta e^{-i\beta} )\rho/\sqrt{2}$, where 
$\theta, \alpha$, and $\beta$ are the three Goldstone fields of the spontaneously broken global 
$SU(2)$  group.   
  In this parameterization  the  Higgs Lagrangian becomes, 
     \begin{equation}
 \frac12 (\partial_{\mu} \rho)^2 \, + \,  \frac{\rho^2}{2}(\partial_{\mu} U^{\dagger}  \partial_{\mu} U) \, - \, {\lambda^2 \over 8} (\rho^2 - v^2)^2 \, .
   \label{higgspart}
   \end{equation}
   where 
  \begin{equation}
  \partial^{\mu} U(x)^{\dagger}  \partial_{\mu} U(x) \, = \, 
   \left[ (\partial_{\mu} \theta)^2  + \cos^2 \theta  (\partial_{\mu} \alpha)^2   +   \sin^2\theta (\partial_{\mu} \beta)^2  \right ] \, . 
  \label{invariant}
  \end{equation}

Integrating-out the Higgs through its equation of motion,  which at low energies becomes an algebraic constraint
  \begin{equation}
  \rho^2 \, = \, v^2 +\frac{2}{\lambda^2} (\partial_{\mu} U^{\dagger }  \partial_{\mu} U) \, ,
   \label{higgseq}
   \end{equation}
   and rescaling,  $U \, \rightarrow  \, v U$, 
we obtain the following effective theory
 \begin{equation}
   \frac{1}{2}(\partial_{\mu} U^{\dagger}  \partial^{\mu} U) \, + \, {1\over 2 \lambda^2 v^4}  (\partial_{\mu} U^{\dagger}  \partial^{\mu} U)^2 \,.
  \label{higgseff}
 \end{equation}
  This is similar to (\ref{eq:goldstone}) with  $\epsilon \, = \, 1$ and   $L_*^4  \, = \, 2/(\lambda^2 v^4)$. 
   The gauge case can be trivially restored by replacing $\partial_{\mu}$ with the covariant  derivatives of the $SU(2)\times U(1)$ group.  
The positive sign of the four-derivative term indicates that the theory can be UV-completed by the Higgs particle.  On the other hand,  for the negative sign no such completion is possible, and the theory chooses the classicalization route.  Thus, by detecting the sign of this operator at low energies we obtain 
the information about which route the theory chooses for its UV-completion. 
 This sign can in principle be read-off from  measuring the sign of the amplitude of longitudinal $WW$-scattering.

\section{Conclusions} 

 The questions we have addressed in this work are: 
  How does a given derivatively-coupled theory chooses  which road towards UV-completion,  Wilsonian versus classicalization, to take? 
  And, what are the low energy observables that determine the road? 
  
   First,  we have seen that the two concepts are inter-exclusive and that  integrating-in weakly coupled Wilsonian physics kills classicalons. 

 Moreover, we have seen that the information about the chosen road 
is encoded in the sign of the derivative couplings, such as, for a 
Goldstone-type particle the quartic coupling in (\ref{eq:goldstone}). 

  Using a simple argument based on consistency of the K\"allen-Lehmann  spectral representation we have shown that such an effective vertex
  with a {\it negative sign}  can never result from integrating-out a sensible weakly-coupled physics. 
   This is in agreement  with the previous arguments \cite{nima} based 
   on superlumnality and dispersion relations.  However, our point is that in case of the negative sign instead of dismissing the theory, the vertex must be treated as {\it fundamental}.   It is precisely for this sign that consistent classicalons appear,  indicating that the theory chooses a non-Wilsonian way of self-UV-completion. 
  In other words, the theory puts up a self-defense by classicalization against  inconsistencies, such as the violation of causality due to superluminality.   
 In this light we have rediscussed the implication of $a$-theorems in the 
 context of non-Wilsonian  physics.        
    
 Applying these ideas to the self-completion of the Standard Model via classicalization we are led to the conclusion that the choice of the road towards  
 UV-completion is encoded in the sign of the scattering amplitude 
 of longitudinal $W$-bosons. 
  Of course, having in mind UV-completion by 
 Higgs or any other  Wilsonian physics, one would never question 
 the positivity of this sign.  But the possibility of alternative UV-completions motivates the check.  Even if a Higgs-like  resonance is discovered at the LHC, 
 to determine the sign would provide a powerful cross-check that we are indeed 
 dealing with the Higgs particle.

\section{Acknowledgements}

It is pleasure to thank Luis Alvarez-Gaume, Siggi Bethke, Daniel Flassig and Alex Pritzel for discussions. 
The work of G.D. was supported in part by Humboldt Foundation under Alexander von Humboldt Professorship,  by European Commission  under 
the ERC advanced grant 226371,  by European Commission  under 
the ERC advanced grant 226371,   by TRR 33 \textquotedblleft The Dark
Universe\textquotedblright\   and  by the NSF grant PHY-0758032. 
The work of C.G. was supported in part by Grants: FPA 2009-07908, CPAN (CSD2007-00042) and HEPHACOS S2009/ESP-1473.

$~~~~$

{\bf Note Added}

$~~~~$

Before submission of this work, a paper \cite{polch} appeared which gives further analysis along the lines of \cite{komar}.

\end{document}